\documentstyle[pra,multicol,aps,amstex]{revtex} 
\begin{document}

\setlength{\textwidth}{150mm}
\setlength{\textheight}{240mm}
\setlength{\parskip}{2mm}

\input{epsf.tex}
\epsfverbosetrue

\draft

\renewcommand{\baselinestretch}{1.0}

\title{Influence of four-wave mixing and walk-off on the self-focusing 
of coupled waves}

\author{L. Berg\'e$^1$, O. Bang$^2$ and W. Krolikowski$^3$}

\address{
$^1$ Commissariat \`a l'Energie Atomique, CEA/Bruy\`eres-le-Ch\^atel, 
B.P. 12, 91680 Bruy\`eres-le-Ch\^atel, France.\\
$^2$ Department of Mathematical Modelling, Technical University of Denmark, DK-2800 Lyngby, Denmark.\\
$^3$ Optical Sciences Centre, Australian National University,
Canberra ACT 0200, Australia.}

\maketitle

\normalsize 
 
\begin{abstract} 
Four-wave mixing and walk-off between two optical beams are investigated for focusing Kerr media. It is shown that four-wave mixing reinforces the self-focusing of mutually-trapped waves by lowering their power threshold for collapse, only when their phase mismatch is small. On the contrary, walk-off inhibits the collapse by detrapping the beams, whose partial centroids experience nonlinear oscillations. 
\end{abstract} 

\pacs{PACS numbers : 42.65.Tg, 42.25.Bs, 05.45.Yv, 42.65.Jx} 

\begin{multicols}{2}

\narrowtext

Self-trapping of intense electromagnetic waves in 
nonlinear media have inspired interest since the earliest days of 
nonlinear optics \cite{chiao}. For a focusing Kerr medium, this process is often described by a (1+1)-dimensional nonlinear Schr\"{o}dinger equation (NLS), when, for instance, optical pulses are assumed to only undergo anomalous group-velocity dispersion (GVD) while their transverse diffraction is neglected. Recent developments based on this model 
rapidly increased due to the possibility of making two wave 
components couple nonlinearly and propagate as mutually-trapped beams \cite{manakov,kang,menyuk}. 
In the seventies, such optical beams with two orthogonal polarized 
components were theoretically predicted to form spatial solitons 
preserving their incident shapes and powers \cite{manakov}. 
However, these solitons were only recently detected in experiments 
\cite{kang}, partly because of the difficulty in overcoming parametric 
four-wave mixing (FWM) processes which become relevant when the two 
orthogonal polarizations are not incoherently coupled \cite{akhmediev}. 
Besides FWM, they moreover undergo linear 
convection, often termed as "walk-off", which can be either attached to the half-difference of their group velocity in birefringent media \cite{menyuk,cao}, or connected with the angle between the transverse and carrier wave vectors in, e.g., biased photorefractive crystals \cite{christodoulides}.

In the previous works, only one-dimensional 
dispersion was considered and the interaction patterns between two partial pulses were analyzed in terms of stable solitons keeping a fixed sech-shape \cite{akhmediev,cao}. However, the underlying assumption that the soliton components preserve their shape while propagating does not hold at higher dimensions. It is indeed well-known that NLS solutions are {\it unstable}, by spreading out or self-focusing until collapse, when transverse diffraction is 
no longer disregarded and enters ($D$+1)-dimensional equations with a number of spatio-temporal dimensions, $D$, at least equal to two \cite{rasmussen}. High-power pulses then sharply increase in amplitude and undergo destructive modulations, which significantly affect their mutual interactions \cite{mackinstrie}. Even if the collapse singularity is ultimately arrested by saturation mechanisms, the dynamics preceding it is important for understanding this growth stage, which 
justifies to study the self-focusing of nonlinear waves when FWM and 
walk-off cannot be ignored. 
In this letter, the influence of both these effects is investigated from ($D$+1)-dimensional coupled 
NLS equations, with emphasis on the potential modifications of the 
power threshold for collapse when $D$=2. 
By means of analytical estimates and numerical verifications, we show 
that FWM contributes to self-focusing by lowering this threshold for pulses with small phase mismatch only, whereas walk-off acts against the collapse by inducing oscillations in the wave trajectories.

We consider the slowly-varying envelopes $u_1({\vec r},z)$ and 
$u_2({\vec r},z)$ of two waves copropagating along the $z$-axis of a 
Kerr medium. Here ${\vec r}$ refers to the vector of a generalized $D$-dimensional transverse diffraction plane, which 
can also include a retarded time variable when anomalous GVD is retained. 
The waves undergo walk-off and FWM as described by the generic model \cite{kang,menyuk,akhmediev,cao}
$$
   i (\partial_z + {\vec \delta_n} \cdot {\vec \nabla}_{\perp}) u_n 
   - \beta_n u_n + {\vec \nabla}_{\perp}^2 u_n 
   + (|u_n|^2 + A |u_{3-n}|^2) u_n
$$
\begin{equation}
   \label{1}
   \displaystyle{+ B u_{3-n}^2 u_n^* = 0, \,\,\,\,\, n = 1,2,}
\end{equation}
where $*$ means complex conjugate, ${\vec \delta}_1 = -{\vec \delta}_2 
= {\vec \delta}$ and $\beta_1 = -\beta_2 = \beta$. In Eq.(1), standard 
notations have been used. In particular, the second term describes walk-off effects, the third term accounts for the mismatch in wave numbers between the components and the last term represents FWM. The constants $A$ and $B$ are positive and they measure the strength of the nonlinear coupling between the two waves. In isotropic media, their values satisfy $A+B$=1 when $u_1$ and $u_2$ represent two orthogonal polarizations of a vector field (as, e.g., $A$=2$B$=2/3 in birefringent fibers \cite{menyuk}), while they are linked by $A$=$2B$=2 for two copropagating beams with scalar amplitudes $u_1$ and $u_2$. For Eq.(\ref{1}) the total power $P = P_{1} + P_{2} = \|u_1\|_2^2 + \|u_2\|_2^2$ is always conserved, and it keeps invariant each individual power $P_n \equiv \|u_n\|_2^2$ when $B = 0$ only. 
For notational convenience, we make use of the standard $L^p$ norms, $\|f\|_p^p 
\equiv \int |f|^p d{\vec r}$, where $\int d{\vec r}$ denotes an 
integration over the $D$-dimensional transverse space. 
Equations (\ref{1}) also conserve the Hamiltonian
$$
   H = \sum_{n=1}^2 \{ \|{\vec \nabla}_{\perp} u_n\|_2^2 
   + \beta_n P_n - \chi_n(A,B)
$$
\begin{equation}
   \label{5}
   \displaystyle{+ \int |u_n|^2 {\vec \delta}_n \cdot 
   {\vec \nabla}_{\perp} \mbox{arg}(u_n) d{\vec r}\},}
\end{equation}
where
$$
   \chi_n(A,B) \equiv \frac{1}{2}\|u_n\|_4^4 
   + \frac{A}{2} \|u_n u_{3-n}\|_2^2 + \frac{B}{2} \mbox{Re} 
   \int (u_{3-n}^2 u_n^{*2}) d{\vec r}
$$
is the nonlinear potential related to the Kerr and FWM contributions. 
Furthermore, we can derive a virial identity describing the evolution 
of the total mean square radius of both waves along $z$. 
This relation governs the second-order $z$-derivative of the integral
\begin{equation}
\label{6}
\displaystyle{{\cal I}(z) \equiv \frac{1}{P} \int |{\vec r} - \langle {\vec r} \rangle|^2 \sum_{n=1}^2 |u_n|^2 d{\vec r},}
\end{equation}
which involves the total center of mass $\langle {\vec r} \rangle \equiv P^{-1} \int 
{\vec r} \sum_{n=1}^2 |u_n|^2 d{\vec r}$. 
After employing the straightforward procedure 
expounded in \cite{mackinstrie}, the virial identity is found 
to read
\begin{equation}
   \label{8}
   \displaystyle{\partial_z^2 {\cal I} = \partial_z^2 I/P - 
   2 |\partial_z \langle {\vec r} \rangle|^2 - 2 \langle {\vec r} \rangle \cdot \partial_z^2 
   \langle {\vec r} \rangle}
\end{equation}
where $I \equiv \int r^2 \sum_{n=1}^2 |u_n|^2 d{\vec r} = P({\cal I} 
+ |\langle {\vec r} \rangle|^2)$,
$$
   \partial_z^2 I = 4DH + 4\sum_{n=1}^2 \{(2-D) 
   \|{\vec \nabla}_{\perp} u_n\|_2^2 - D\beta_n P_n
$$
\begin{equation}
   \label{7}
   \displaystyle{+ \frac{\delta_n^2 P_n}{2} - B {\vec \delta}_n \cdot 
   \mbox{Im} \int {\vec r} (u_{3-n}^2 u_n^{*2}) d{\vec r} \},}
\end{equation}
\begin{equation}
   \label{9}
   \displaystyle{P \partial_z \langle {\vec r} \rangle = {\vec M}^{\delta} \equiv 
   {\vec M}^0 + \sum_{n=1}^2 {\vec \delta}_n P_n,}
\end{equation}
\begin{equation}
   \label{10}
   \displaystyle{{\vec M}^0 = 2 \mbox{Im} \sum_{n=1}^2 \int (u_n^* 
   {\vec \nabla}_{\perp} u_n) d{\vec r},}
\end{equation}
with $\delta_n^2 = |{\vec \delta}_n|^2$. Here, ${\vec M}^0$ denotes the conserved total momentum. When collapse develops, the integral ${\cal I}(z)$ vanishes at a finite propagation distance $z_c$, implying ${\cal I}_n \equiv P_n^{-1} \|({\vec r} - \langle {\vec r} \rangle) u_n\|_2^2 \rightarrow 0$ as $z \rightarrow z_c$. In this limit, the gradient norms of $u_n$ must blow up by virtue of the inequality $P_n \leq (4/D^2) {\cal I}_n \|{\vec \nabla}_{\perp} u_n\|_2^2$, which makes the wave amplitude diverge in turn. In NLS systems, such a blow-up generally takes place at propagation distances shorter than $z_c$ \cite{rasmussen}, and it arises when the total power exceeds some threshold. For instance, when considering a single wave governed by the (2+1)-dimensional NLS equation $i\partial_z u+{\vec \nabla}_{\perp}^2 u+|u|^2u=0$, blow-up can only develop if $P$ exceeds the threshold value $P_c \simeq 11.7$, which is close to the critical power $4 \pi$ fixing $H=0$ for Gaussian beams \cite{mackinstrie,weinstein}. In the present context, we determine the influence of FWM and walk-off on self-focusing by dividing the analysis into two separate parts devoted to FWM alone and to walk-off alone.

{\em 1. FWM without walk-off :} For the case $B \neq 0$ and ${\vec \delta} = {\vec 0}$, the virial relation (\ref{8}) can be bounded from above when $D \geq 2$ as follows 
\begin{equation}
\label{13}
\displaystyle{\partial_z^2 {\cal I} \leq 4D(H - \beta P_1 + \beta P_2)/P - 2 |{\vec M}^0|^2/P^2,}
\end{equation}
where the strict equality applies to the case $D = 2$ only. Thus, a total collapse of the wavefunction, in the sense ${\cal I}(z) \rightarrow 0$, always occurs for $D \geq 2$, whenever $(H + |\beta|P)$ is negative. In (2+1) dimensions, the threshold power for collapse can be estimated by bounding the Hamiltonian from below with a combination of Cauchy-Schwarz and Sobolev [$\|u_n\|_4^4 \leq (2/P_c)\|{\vec \nabla}_{\perp} u_n\|_2^2 \|u_n\|_2^2$] inequalities :
\begin{equation}
   \label{15}
   H + |\beta|P \geq \sum_{n= 1}^2 [1 - \frac{P_n}{{\cal P}_c^{B}}] 
   \|{\vec \nabla}_{\perp} u_n\|_2^2.
\end{equation}
Since the left-hand side of Eq.(\ref{15}) is finite, a wave collapse with $\|\vec{\nabla}_\perp u_n\|_2^2 \rightarrow +\infty$ cannot occur when the individual power of both components is below the threshold
\begin{equation}
   \label{15b}
   {\cal P}_c^{B} \equiv P_c/(1+A+B).
\end{equation}
Hence, $P_n<{\cal P}_c^{B}$ is a sufficient condition for no collapse. In terms of the conserved total power $P$, collapse is surely avoided when $P$ is below ${\cal P}_c^{B}$, but due to power transfers between both components, the bound $P<{\cal P}_c^{B}$ can significantly underestimate the actual threshold for collapse when $B\ne0$, as was observed in similar studies of quadratic nonlinear media \cite{Bang97oc}. In spite of this, when the two components are initially identical, Eq.(\ref{1}) with $\beta = 0$ reduces by symmetry to the single-wave NLS equation $i\partial_z u + {\vec \nabla}_{\perp}^2 u + (1+A+B) |u|^2u=0$ and power transfers are eliminated. ${\cal P}_c^{B}$ thus restores the exact threshold power for collapse of each component and, therefore, $P > 2{\cal P}_c^{B}$ can here yield a reasonably-good estimate of the exact collapse threshold for small mismatch values, provided that $u_1(\vec{r},0)\simeq u_2(\vec{r},0)$. In this situation, Eq.(\ref{15b}) shows that FMW lowers the power threshold for collapse and strengthens the self-focusing dynamics, compared with that of incoherently-coupled waves ($B = 0$) \cite{mackinstrie}.

Using an iterative, radially-symmetric, mid-point Crank-Nicholson
finite-difference scheme, we have performed numerical simulations
of Eq.(\ref{1}) in (2+1) dimensions with the Gaussian initial 
condition
\begin{equation}
   \label{16bis}
   u_n({\vec r},0) = \sqrt{P_n/\pi\Delta^2} \exp{(-r^2/2\Delta^2)}.
\end{equation}
We used a resolution of $\Delta r = 10^{-3}$ over the interval $r$=$[-20,+20]$ and a stepsize of $\Delta z = 10^{-3}$, which kept the relative deviation of $P$ from its initial value below $10^{-4}$ over a distance $z$=[0,50].

\begin{figure}
\setlength{\epsfxsize}{8.0cm}
\epsffile{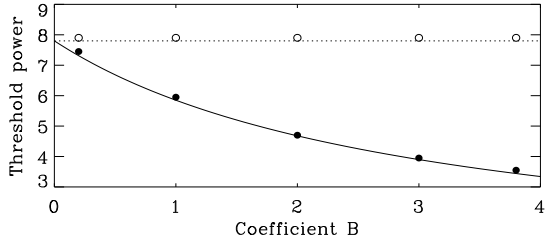}
\vspace{2mm}
\caption{Collapse threshold power $2{\cal P}_c^B$ versus $B$ for $u_1(\vec{r},0)$=$u_2(\vec{r},0)$ with $A$=2 (solid line). The dotted line represents $2{\cal P}_c^0$. The black and white marked points indicate the numerically-calculated collapse thresholds for $\beta = 0$ and $\beta = 20$, respectively.}
\end{figure}

The solid line in Figure 1 shows the collapse threshold $2{\cal P}_c^B$ versus the FWM coefficient $B$ when $P_1 = P_2 = P/2$ and $\Delta = 2$. For $\beta$=0, this threshold power agrees with its numerical counterpart, since the dynamics is entirely determined by symmetry from the single-wave NLS equation. For $\beta\ne0$, this symmetry reduction does no longer apply and the waves exchange power. However, at large mismatches ($\beta \gg 1$), the FWM terms can merely be averaged out ($B \rightarrow 0$) from Eq.(\ref{1}) after applying the phase transform $u_n \rightarrow u_n \exp{(-i \beta_n z)}$, so that power transfers disappear and the collapse threshold is simply given by $2 {\cal P}_c^0$, as confirmed by the white marked points of Fig. 1. In contrast, a more intriguing behavior appears for moderate values of $\beta$. For $\beta$ ranking as $1 < \beta < 10$, the two components were indeed observed to diffract with no significant power exchange in the range $P < 2 {\cal P}_c^0$. For $P$ above $2{\cal P}_c^0$, they started to periodically transfer their power between each other, which generated strong resonances following several focusing/defocusing cycles. Rapid oscillations in each partial power created spiky resurgences in the wave dynamics, which made it computationally difficult to determine an accurate collapse threshold. Despite this, a non-zero $\beta$ always counteracts the self-focusing promoted at exact phase matching $(\beta = 0)$, by increasing the power threshold for collapse.

{\em 2. Walk-off without FWM :} 
We now consider the opposite situation ${\vec \delta} \neq {\vec 0}$ 
and $B = 0$. 
As the nonlinearities only depend on intensities, the phase mismatch 
plays no role in the collapse dynamics and we henceforth set $\beta_n = 0$ in Eq.(\ref{1}). In this case, equations (\ref{1}) preserve the individual powers $P_n$ and, for $D \geq 
2$, the virial integral (\ref{8}) satisfies
\begin{equation}
   \label{22}
   \displaystyle{\partial_z^2 {\cal I} \leq \frac{8}{P} H 
   + 2 \delta^2 - \frac{2}{P^2}|{\vec M}^{\delta}|^2,}
\end{equation}
where the strict equality still applies to $D = 2$ only. From Eq.(\ref{22}), we infer that collapse inevitably develops when $H + \delta^2 P/4 \leq 0$. This requirement reflects the interplay between the focusing nonlinearities in the invariant $H$ and the increase in $\delta^2$ of the total mean-square radius due to walk-off. As a result, the finite distance $z_c^{\delta}$ at which ${\cal I}(z)$ tends to 
vanish in case of collapse is larger than the collapse focus $z_c^0$ 
corresponding to ${\vec \delta} = {\vec 0}$. Therefore, the collapse dynamics should be delayed and possibly arrested when ${\vec \delta} \neq {\vec 0}$. Let us check this property with the $(2+1)$-dimensional Gaussian beams (\ref{16bis}). Without walk-off, these beams possess the Hamiltonian $H = (P/\Delta^2)(1 - P/{\cal P}_{th})$ with the collapse threshold
\begin{equation}
   \label{23}
   \displaystyle{{\cal P}_{th} = \frac{4\pi(1 + \rho)^2}{1 + \rho^2 
   + 2 \rho A},}
\end{equation}
where $\rho \equiv P_1/P_2$. In the presence of walk-off, the estimate (\ref{22}) yields 
the effective critical power for collapse
\begin{equation}
   \label{23bis}
   \displaystyle{{\cal P}_c^{\delta} = {\cal P}_{th}\{1 + 
   \frac{\delta^2 \Delta^2 \rho}{(1+\rho)^2}\},}
\end{equation}
which increases with the walk-off length $\delta$. Figure 2 shows this critical power for equal waves with $P_1 = P_2$. The threshold (\ref{23bis}) is associated with a total collapse taking place at the center $\langle {\vec r} \rangle = {\vec 0}$ and it raises with $\delta$, which is confirmed numerically. At a given power, walk-off is thus able to delay and even arrest such a collapse, as it can be guessed from the virial expression ${\cal I}(z) = (4z^2/\Delta^2 {\cal P}_{th})({\cal P}_c^{\delta} - P) + \Delta^2$.

\begin{figure}
\epsfxsize 8cm
\epsffile{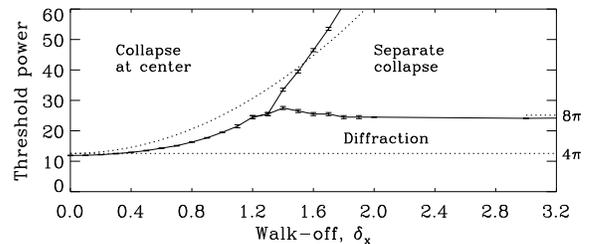}
\caption{Collapse threshold versus $\delta \equiv \delta_x (\delta_y = 0)$ for equal waves with $A$=1. The solid curve, where error bars have been specified for the sake of accuracy, represents the numerical result for $P_1$=$P_2$ and $\Delta$=2. The dotted lines correspond to the analytical predictions ${\cal P}_{th}(1+\delta^2), 8\pi$, and ${\cal P}_{th}$=$4\pi$.}
\label{fig2}
\end{figure}

The agreement between the theoretical estimate (\ref{23bis}) and the numerical simulations is reasonably good for small $\delta$ values. For large $\delta$, Figure 3 displays that walk-off significantly competes with the collapse by modifying the trajectories of both components in the transverse plane and by detrapping them. The centroids of the components are then shifted from $\langle {\vec r} \rangle = {\vec 0}$ and further fuse again at the origin, where collapse occurs [Fig.~3(a)]. This explains the discrepancies in Fig. 2 between the true collapse threshold and $P_c^{\delta}$ for a total collapse on the center. Alternatively, when the two waves contain enough power for promoting individual collapses $(P_1 = P_2 > 4 \pi)$, walk-off can make them separate from each other and collapse on their own center of mass, far away from the origin [Fig.~3(b)]. In this case, Eq.(\ref{22}) emphasizes that walk-off becomes important whenever $|H| \ll \delta^2P/4$, which yields a bound for $\delta$, namely $\delta > (2/\Delta)\sqrt{P/{\cal P}_{th} - 1}$, above which the two components can collapse separately with individual mean-square radii tending to zero. This results in a saturation plateau at $P = 8\pi$, towards which the collapse threshold saturates from $\delta > 1$. Conversely, in the diffraction regime where collapse does not occur, the two components can undergo one or more crossings before decaying into noise, or continue to walk off away from each other while they both decouple and disperse.

\begin{figure}
\setlength{\epsfxsize}{8.0cm}
\epsffile{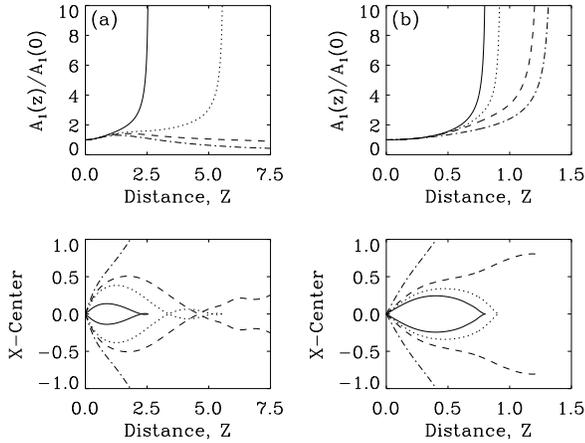}
\vspace{2mm}
\caption{Dynamics of coupled waves undergoing walk-off for (a) $P$=16 with $\delta$=0.3 (solid), $\delta$=0.7 (dotted), $\delta$=0.8 (dashed), $\delta$=1.0 (dash-dotted), and (b) $P = 40$ with $\delta$= 1.0 (solid), $\delta$=1.3 (dotted), $\delta$=1.6 (dashed), $\delta$=3.0 (dash-dotted).}
\end{figure}

To understand the motion of the individual centers of mass, $\langle {\vec r}_n (z) \rangle \equiv P_n^{-1} \int {\vec r} |u_n|^2 d{\vec r}$, 
we multiply Eq.(\ref{1}) by $({\vec r} u_n^*)$ and $({\vec \nabla}_{
\perp} u_n^*)$ and combine the results to derive
\begin{equation}
   \label{DS1}
   \displaystyle{\frac{d^2}{dz^2} \langle {\vec r}_n \rangle + \frac{2 A}{P_n} \int |u_{3-n}|^2 {\vec \nabla}_{\perp} |u_n|^2 d{\vec r} = 0,}
\end{equation}
where $d_z \langle {\vec r}_n \rangle (0) = {\vec \delta}_n$. 
The integral in (\ref{DS1}) measures the flux induced on the $n$th 
wave by its neighbour, which can make $\langle {\vec r}_n \rangle$ periodically displace to a fixed point $\langle {\vec r}_n^{\,0} \rangle$, then go back to the center. For equal components modelled by Gaussians $|u_n|^2 = (P/2\pi\Delta^2) \mbox{e}^{-|{\vec r} - \langle {\vec r}_n \rangle|^2/\Delta^2}$ located symmetrically from the origin, Eq.(\ref{DS1}) reduces to $d_z^2 \langle {\vec r}_n \rangle + (AP/\pi \Delta^4) \langle {\vec r}_n \rangle \mbox{e}^{-2 \langle r_n \rangle^2/\Delta^2} = 0$ and $\langle {\vec r}_n \rangle$ has indeed an harmonic motion, which is damped in collapse regimes [$\Delta(z) \rightarrow 0$] and relaxes otherwise [$\Delta(z) \rightarrow + \infty$]. Along one arch of oscillation, we can moreover suppose that the waves keep a fixed width $\Delta$, so that the same equation provides the relation linking $\delta_n$ to the maximal displacement of the beam centroid :
$\langle r_n^0 \rangle = \frac{\Delta}{\sqrt{2}} \sqrt{\ln{(1 - 2 \pi \Delta^2 \delta_n^2/AP)^{-1}}}$. Hence, the constraint that oscillations develop only if $\langle r_n^0 \rangle$ remains finite imposes a critical value to the walk-off length, namely $\delta_n^2 \leq \delta_c^2 = AP/2 \pi \Delta^2$. This critical value, below which both waves are reflected back to the origin, can also be estimated by considering two beams launched in the medium at angles $\theta_0$ and $-\theta_0$ with respect to {\em z}, such as $\delta_n = 2 n_0 \tan\theta_0$. By following the ray trajectory associated with one beam, geometrical optics predicts that this ray experiences a total internal reflection if $\theta_0$ satisfies $\cos\theta_0=n_0/(n_0+n(0))$, where $n(0)$ and $n_0$ denote the values of the refractive index on the axis and at infinity, respectively. This relation then expresses as $\tan^2\theta_0= I /n_0^2$, where $I$ is the light intensity at center. For two beams symmetrically displaced from the origin, $I$ is mainly given by the interaction part of $H$, i.e., $H_{int}/P \sim AP/8\pi \Delta^2$, which thus restores $\delta_c$.

In conclusion, we have shown that in Kerr media FWM lowers the threshold power for collapse of two coupled waves, when they are close to exact phase matching only. The resulting enhancement of self-focusing, however, ceases at moderate mismatch values, for which the waves undergo sharp resonances. On the other hand, walk-off delays and may stop the self-focusing by altering the beam trajectories. For comparison, two weakly-separated (1+1)-dimensional solitons in birefringent fibers are known \cite{cao} to either bounce back and forth while forming a trapped state, or separate by developing nonlinear oscillations, whenever their amplitude is above or below a threshold depending nonlinearly on the walk-off length. At higher dimension, analogous behaviors also characterize high-power pulses, whose components can either mutually attract or separate with walk-off. However, the self-focusing dynamics here offers a richer variety of interaction regimes, among which, e.g., two coupled components, being even initially superimposed, can split before fusing into a single lobe or reversely collapse after splitting. Finally, our analysis suggests that, in saturable optical media, high-dimensional beams should still be capable of coalescing into a robust light bullet, which may be desirable for optical switching applications.

\end{multicols}
 
\end{document}